\documentclass[preprint,a4paper,superscriptaddress,floatfix]{revtex4-2}
\usepackage{graphicx,xcolor}
\usepackage{amsmath,amsfonts,amssymb,bm}

\usepackage{listings,soul}

\usepackage[
unicode=true,
bookmarks=true,%
bookmarksnumbered=true,%
hyperfootnotes=false,
colorlinks=true,%
linkcolor=blue,%
citecolor=blue,%
urlcolor=blue
]{hyperref}

\newenvironment{mat}{\left[\begin{array}{ccccccccccccccc}}{\end{array}\right]}
\newcommand\bcm{\begin{mat}}
\newcommand\ecm{\end{mat}}

\newenvironment{cmat}{\left(\begin{array}{ccccccccccccccc}}{\end{array}\right)}
\newcommand\bcrm{\begin{cmat}}
\newcommand\ecrm{\end{cmat}}

\newenvironment{rmat}{\left[\begin{array}{rrrrrrrrrrrrr}}{\end{array}\right]}
\newcommand\brm{\begin{rmat}}
\newcommand\erm{\end{rmat}}

\definecolor{red1}{HTML}{E24647}
\definecolor{blue1}{RGB}{160, 210, 250}
\definecolor{pink1}{HTML}{ffbbbb}
\definecolor{green1}{RGB}{0, 125, 0}
\definecolor{green2}{RGB}{150, 220, 150}
\definecolor{orange1}{HTML}{faaa82}
\definecolor{orange2}{HTML}{ff6f00}
\definecolor{purple1}{HTML}{deadff}
\definecolor{purple2}{HTML}{9c00ff}
\definecolor{sicolor}{HTML}{b0d9c6}

\begin{document}

%TC:ignore

% \title{Topological state transfer by twisting Kresling origami\ym{?}}
\title{Topological state transfer in Kresling origami}
\author{Yasuhiro Miyazawa}
\affiliation{Department of Aeronautics and Astronautics, University of Washington, Seattle, Washington 98195, USA}

\author{Chun-Wei Chen}
\affiliation{Department of Aeronautics and Astronautics, University of Washington, Seattle, Washington 98195, USA}

\author{Rajesh Chaunsali}
\affiliation{Department of Aerospace Engineering, Indian Institute of Science, Bangalore 560012, India}

\author{Timothy S. Gormley}
\affiliation{Department of Aeronautics and Astronautics, University of Washington, Seattle, Washington 98195, USA}

\author{Gloria Yin}
\affiliation{Department of Aeronautics and Astronautics, University of Washington, Seattle, Washington 98195, USA}

\author{Georgios Theocharis}
\affiliation{Laboratoire d'Acoustique de l'Universit\'e du Mans (LAUM), UMR 6613, Institut d'Acoustique - Graduate School (IA-GS), CNRS, Le Mans Université, France}

\author{Jinkyu Yang}
% \email[]{jkyang@aa.washington.edu}
\affiliation{Department of Aeronautics and Astronautics, University of Washington, Seattle, Washington 98195, USA}

\date{\today}

\begin{abstract}

Topological mechanical metamaterials have been widely explored for their boundary states, which can be 
robustly isolated or transported in a controlled manner. However, such systems often require pre-configured design or complex active actuation for wave manipulation. 
Here, we present the possibility of in-situ transfer of topological boundary modes by leveraging the reconfigurability intrinsic in twisted origami lattices.
In particular, we employ a dimer Kresling origami system consisting of unit cells with opposite chirality, which couples longitudinal and rotational degrees of freedom in elastic waves.
The quasi-static twist imposed on the lattice alters the strain landscape of the lattice, thus significantly affecting the wave dispersion relations and the topology of the underling bands.
This in turn facilitates an efficient topological state transfer from one edge to the other.
This simple and practical approach of energy transfer in origami-inspired lattices can thus inspire a new class of efficient energy manipulation devices.

\end{abstract}

\maketitle
%TC:endignore

\section{Introduction}

In recent decades, mechanical metamaterials have been considered an ideal medium to design effective wave manipulation systems, leveraging their high degree of freedom in design and the  tailorability of their mechanical properties~\cite{Hussein2014,Bertoldi2017,Surjadi2019}.
Mechanical metamaterials are primarily highlighted for their unconventional dynamic responses that can offer rich applications, including vibration isolation~\cite{Miniaci2016,Krushynska2017,Li2017,Wu2020,Liu2020,Zhang2021}, wave guiding~\cite{Babaee2016,Bilal2017,Li2019,Jin2020}, and energy harvesting~\cite{Li2017,Yang2019,Deng2021}.
More recently, the discovery of the topological insulators has significantly influenced and extended the design of wave-guiding mechanical metamaterials~\cite{Roman2015,Wang2015}.
Topological insulators are known for their robustness of the boundary states~\cite{Hasan2010}, which can lead to interesting wave properties for classical systems~\cite{Ozawa2019, Ma2019}.
In addition, inspired by the concept of Thouless adiabatic quantized pumping~\cite{Thouless1983}, which is based on an adiabatic cyclic modulation of the one-dimensional (1D) potential parameters,
diverse lattice models have demonstrated the efficient transfer of the energy between the states, by employing the spatially~\cite{Kraus2012,Verbin2015,Zilberberg2018,Rosa2019,Chen2021a} or temporally~\cite{Chaunsali2016,Oudich2019,HotzenGrinberg,Xu2020a,Xia2021a} varying configurations.

In parallel, recent studies have explored protocols of exploiting the topological properties of several systems, quantum~\cite{Mei2018,Longhi2019,Longhi2019a,Estarellas2017,Lang2017} and classical~\cite{Brouzos2020}, for the state transfer of localized states, a process of great importance for quantum technologies. 
The key advantage topology offers in such process is the inherent protection of the topologically protected boundary against disorder.
This is a significant improvement over the conventional systems, which usually involves energy leakage due to the noises and fabrication errors in the system. 
However, even in topological systems, most systems suffer from the inefficiency in energy management.
They either use pre-configured passive lattices that are not tunable after assembly~\cite{Kraus2012,Verbin2015,Zilberberg2018,Rosa2019,Chen2021a}, or use a complex setting of active elements spanning the entire structure~\cite{HotzenGrinberg,Xia2021a}, which is cumbersome for practical purposes.

To address such limitations, we employ a mechanical lattice consisting of origami-based mechanical units, which offers a simple strategy to tune wave dispersion relationships in-situ and facilitate a robust state transfer.
Origami has served as an efficient design principle to tailor kinematic and static responses in mechanical metamaterials.
Lately, the origami-inspired metamaterials have also been found to be effective in realizing the wave manipulation in them, leveraging their high reconfigurability and controllability~\cite{Thota2018,Yasuda2019}.

In this study, we specifically consider the 1D dimer lattice composed of the origami unit cells to explore the tunable wave dispersion relationship and the emerging topological edge states.
The origami unit is based on the Kresling pattern with opposite chirality, which shows the axial-rotation coupling and nonlinear static responses~\cite{Kresling2012, Yasuda2017a}.
Interestingly, due to this axial-rotation coupling in our system, alternating chirality along the length of the lattice itself leads to the opening of a lower band gap.
We find that this lower band gap is topologically nontrivial, and we demonstrate experimentally that this lower band gap hosts topologically protected edge states.

More interesting mechanism is obtained when the lattice is twisted, which incurs the change of the axial strain and effective stiffness landscape along the lattice.
As a result, the linear wave dispersion relationships of the system are altered, and in this process, we witness the emergence of another band gap in higher frequency regime. 
We report that this upper band gap transitions from topologically trivial to nontrivial states by changing the twist angle in time, and therefore, facilitates a robust state transfer in the lattice from one side of the lattice to the other.
We numerically show such efficient boundary state transfer in the higher frequency regime and evaluate the transfer fidelity.

Notably, we find that such state transfer observed in the upper band gap cannot be achieved in the lower band gap, since it preserves its topologically nontrivial characteristics at any twist angle.
This implies that by leveraging the coupled dynamics in our origami system, we can realize very distinctive energy management capabilities in different frequency regimes.
Thus, our origami system hints an efficient and controllable way to manipulate multiple wave phenomena hosted within the single topological mechanical metamaterial by combining the concepts of topology and origami.

\section{Results}

\subsection{Physical set-up and mathematical model}\label{setup}
\begin{figure}[htbp]
    \centering
    \includegraphics[width=0.5\linewidth]{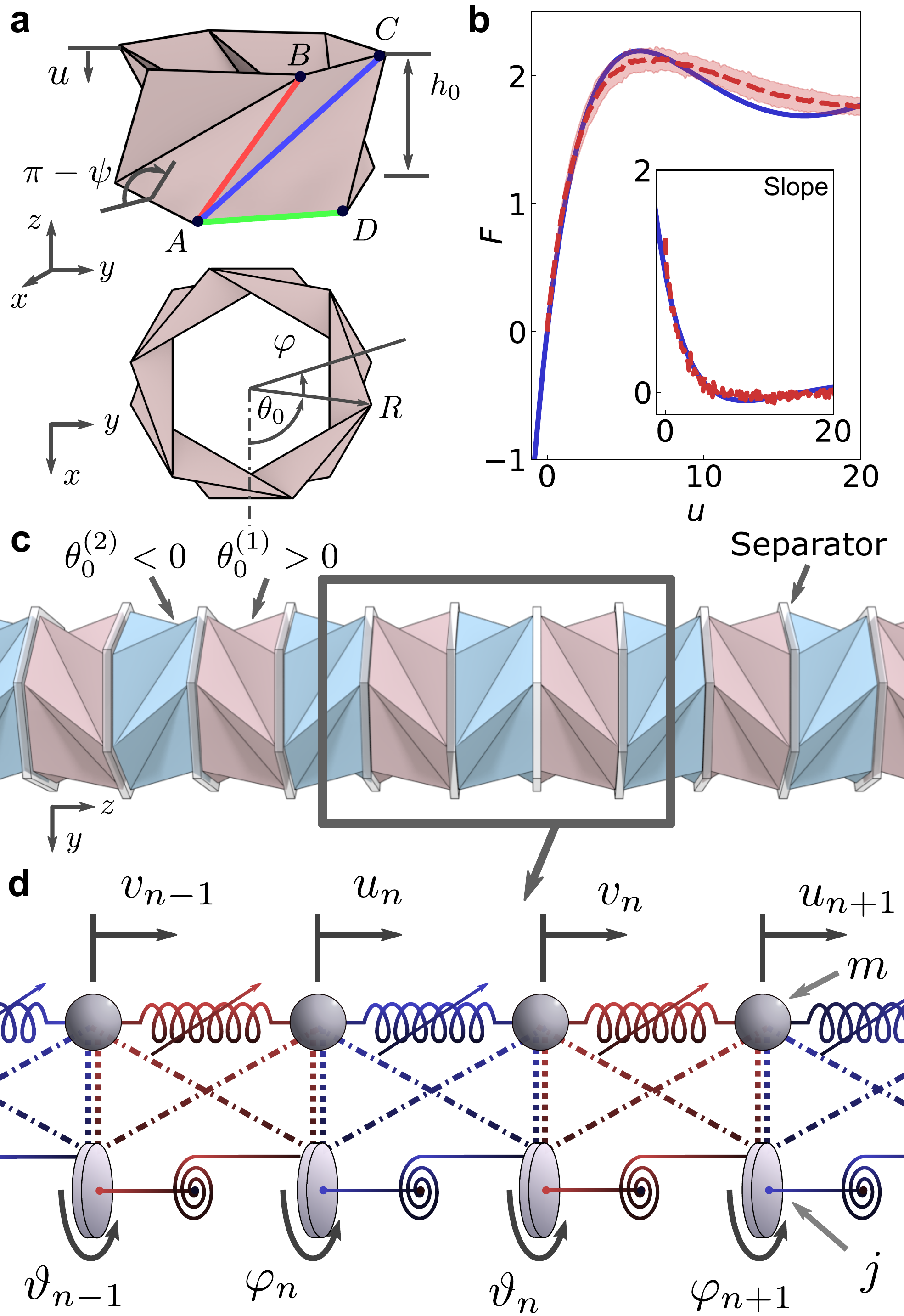}
    \caption{\textbf{The schematic illustrations of the system.}
    \textbf{a} The unit cell depicted in side and top view with geometrical parameters and variables.
    \textbf{b} The force-displacement relationship of the unit cell under pure axial compression along $z$-axis is shown for the geometrical parameter $h_0=30$~mm, $\theta_0=70^\circ$, and $R=36$~mm.
    Red dashed line, average force of experimental result from 18 unit cells; red shaded area, standard deviation; blue solid line, truss model curve.
    The inset figure shows the slope of averaged experimental and analytical force displacement curves.
    \textbf{c} 1D dimer Kresling lattice with polygonal separators.
    Red-colored unit cells have positive chirality $\theta_0^{(1)}>0$, and the blue-colored unit cells have negative chirality $\theta_0^{(2)}<0$.
    \textbf{d} Coupled 1-dimensional phononic lattice as a model of Kresling lattice.
    Polygonal separators are modeled as lumped mass $m$ and disc $j$.
    Lumped masses are connected with nonlinear springs, and so are the discs.
    Adjacent masses and discs are connected with nonlinear springs as well, denoted as dashed and dash-dotted lines.
    }
    \label{fig:schematic}
\end{figure}

We employ the geometrical and kinematic parameters shown in Fig.~\ref{fig:schematic}a to describe the Kresling-patterned unit cell~\cite{Kresling2012}.
The unit cell has two degrees of freedom: translation along $z$-axis~($u$) and rotation about $z$-axis~($\varphi$).
As the unit cell gets compressed~(i.e., $u$ changes), the top surface rotates ($\varphi$ varies), therefore exhibiting coupled behavior.
The resultant force-displacement relationships under pure axial compression~(i.e., no external torque applied) in the experiment is shown in Fig.~\ref{fig:schematic}b, denoted as a red dashed line.
% With the chosen geometrical parameters~($h_0=30$~mm, $\theta_0=70^\circ$, and $R=36$~mm as shown in Fig.~\ref{fig:schematic}a), the unit cell exhibits strain-softening behavior.

To model this coupled folding behavior, we employ a truss model~\cite{Yasuda2019} by replacing the creases along segments $\overline{AB}$ and $\overline{AC}$~(denoted by red and blue solid lines in Fig.~\ref{fig:schematic}a) with linear spring elements, and segment $\overline{AD}$~(green solid line in Fig.~\ref{fig:schematic}a) with a linear torsion spring.
The model gives the total potential energy expressed as,
\begin{align} \label{eq:potential}
    \Pi=\frac{N_p}{2}\left[k_a(a-a_0)^2+k_b(b-b_0)^2+2k_\psi(\psi-\psi_0)^2\right]-Fu-T\varphi,
\end{align}
where $N_p$ is the number of vertices of the polygonal cross-section, $k_a$ and $k_b$ are the linear spring coefficients of the truss elements along $\overline{AB}$ and $\overline{AC}$ respectively, and $k_\psi$ is the torsion spring coefficient of the element along $\overline{AD}$.
The $a$ and $b$ are the lengths of the element $\overline{AB}$ and $\overline{AC}$, and $\psi$ is the angle between the horizontal surface and the triangular facet (e.g., $\triangle ACD$), which are function of $u$ and $\varphi$~(see {Supplementary Note 1} for the explicit expressions).
The subscripted values $a_0$, $b_0$, and $\psi_0$ correspond to their initial lengths and angle.
$F$ is the axial force along $z$-axis, and $T$ is the torque about $z$-axis.

By applying the principle of minimum potential energy, we obtain $F$ and $T$ as a function of $u$ and $\varphi$~(see {Supplementary Note 1}).
This analytical force-displacement relationship is shown in Fig.~\ref{fig:schematic}b, denoted as blue solid line.
Here, the spring coefficients are empirically determined using the least-square method to fit the model curve with the experimental curve.
As a result, we can see the model agrees well with the experimental force-displacement one.
Note that the slopes of the experimental and analytical force-displacement curves---representing the stiffness of the system---are also in agreement as shown in the inset figure of Fig.~\ref{fig:schematic}b.

Having the mathematical model for the unit cell, we now consider the 1D dimer lattice consisting of two different types of unit cells: one with positive chirality $\theta_0^{(1)}>0$ and the other with negative chirality $\theta_0^{(2)}<0$.
Figure~\ref{fig:schematic}c shows the dimer lattice, where positive chirality unit cells~(red-colored) and negative chirality unit cells~(blue-colored) are connected through the polygonal separator, which has mass $m$ and rotational inertia $j$ about $z$-axis.

If Kresling unit cells serve as inter-polygonal springs while having negligible mass and inertia compared to the separators (see {Supplementary Table 1}),
we obtain the equations of motion of the dimer Kresling lattice,
\begin{subequations}\label{eq:eom}
\begin{align}
    m\Ddot{u}_n&+F_2\left(u_n-v_n,\varphi_n-\vartheta_n\right)-F_1\left(v_{n-1}-u_n,\vartheta_{n-1}-\varphi_n\right)=0,\\
    m\Ddot{v}_n&-F_2\left(u_n-v_n,\varphi_n-\vartheta_n\right)+F_1\left(v_n-u_{n+1},\vartheta_n-\varphi_{n+1}\right)=0,\\
    j\Ddot{\varphi}_n&+T_2\left(u_n-v_n,\varphi_n-\vartheta_n\right)-T_1\left(v_{n-1}-u_n,\vartheta_{n-1}-\varphi_n\right)=0,\\
    j\Ddot{\vartheta}_n&-T_2\left(u_n-v_n,\varphi_n-\vartheta_n\right)+T_1\left(v_n-u_{n+1},\vartheta_n-\varphi_{n+1}\right)=0,
\end{align}
\end{subequations}
where $u_n,~v_n$ and $\varphi_n,~\vartheta_n$ are axial displacement and rotational angle of the odd- and even-numbered polygonal separators respectively.
The subscripts $1$ and $2$ of the force and torque functions correspond to the positive and negative chirality unit cells.
%Note that in the above equations of motion, we assumed the same mass and rotational inertia for the odd- and even-numbered polygonal separators.
Figure~\ref{fig:schematic}d shows a schematic of the mathematical model of this coupled system, where the mass and rotational inertia are considered separately, such that the two lattices are connected to each other with nonlinear springs to represent the coupled nature.

\subsection{Tunable wave dispersion relationship}\label{dispersion}
From now on, we consider the dimer Kresling lattice with opposite and equal magnitude chirality, namely odd- and even-numbered Kresling unit cells exhibit $\theta_0^{(1)}=|\theta_0|$ and $\theta_0^{(2)}=-|\theta_0|$.
For the rest of the geometrical parameter of the unit cells, we take them being equal~($h_0^{(1)}=h_0^{(2)}=h_0$ and $R^{(1)}=R^{(2)}=R$).
Our interest is the dynamics of small amplitude elastic waves, and therefore, we start by linearizing Eq.~\eqref{eq:eom}:
\begin{subequations}\label{eq:eom_linear}
\begin{align}
    m\Ddot{u}_n&+\beta_{11}(u_n-v_n)-\alpha_{11}(v_{n-1}-u_n)+\beta_{12}(\varphi_n-\vartheta_n)-\alpha_{12}(\vartheta_{n-1}-\varphi_n)=0,\\
    m\Ddot{v}_n&-\beta_{11}(u_n-v_n)+\alpha_{11}(v_{n}-u_{n+1})-\beta_{12}(\varphi_n-\vartheta_n)+\alpha_{12}(\vartheta_{n}-\varphi_{n+1})=0,\\
    j\Ddot{\varphi}_n&+\beta_{21}(u_n-v_n)-\alpha_{21}(v_{n-1}-u_n)+\beta_{22}(\varphi_n-\vartheta_n)-\alpha_{22}(\vartheta_{n-1}-\varphi_n)=0,\\
    j\Ddot{\vartheta}_n&-\beta_{21}(u_n-v_n)+\alpha_{21}(v_{n}-u_{n+1})-\beta_{22}(\varphi_n-\vartheta_n)+\alpha_{22}(\vartheta_{n}-\varphi_{n+1})=0,
\end{align}
\end{subequations}
where $\alpha$ and $\beta$ are the linear coefficients of the positive and negative chirality unit cells, respectively~(see {Supplementary Note 2} for the detail).
Furthermore, we substitute the Bloch wave solution in Eq.~\eqref{eq:eom_linear} and perform the Fourier transformation to get the eigenvalue problem,
\begin{align}\label{eq:linear_ansatz}
     \omega^2_{k} \bm{u}_k = \hat{D}_k \bm{u}_k
\end{align}
where $k$ and $\omega$ are the wave number and angular frequency, respectively; $\hat{D}_k$ is the dynamical matrix and $\bm{u}_k=(\sqrt{m}u_k, \sqrt{m}v_k, \sqrt{j}\varphi_k, \sqrt{j}\vartheta_k)^T$ is the eigenvector.
Solution to this eigenvalue problem gives the wave dispersion relationship $\omega(k)$ vs. $k$.

\begin{figure}[htbp]
    \centering
    \includegraphics[width=0.5\linewidth]{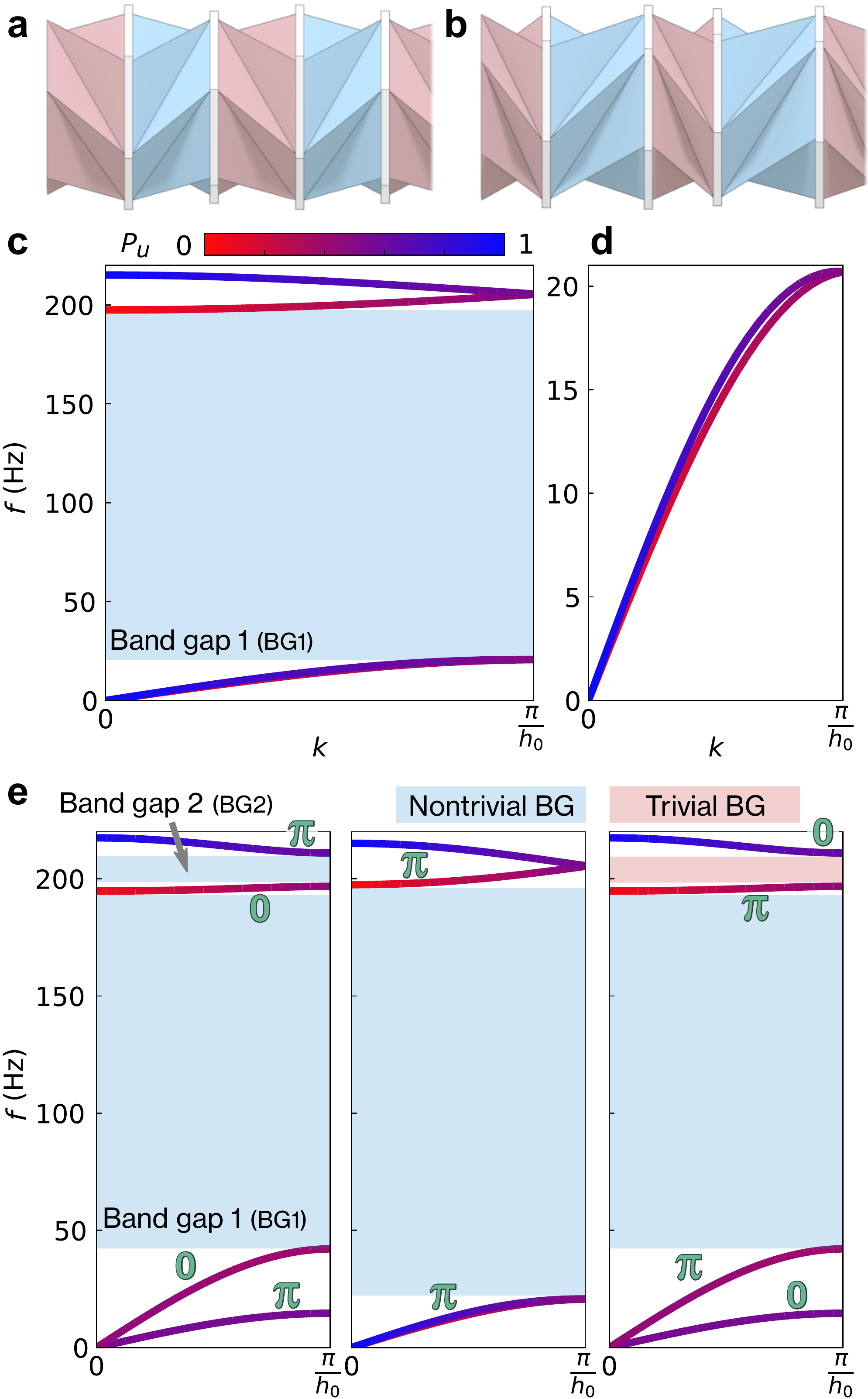}
    \caption{
    \textbf{The dimer Kresling lattices and wave dispersion relationships.}
    The dimer Kresling lattice \textbf{a} under natural condition and \textbf{b} with supercells quasi-statically rotated for $25^\circ$.
    Wave dispersion relationships of the dimer Kresling lattice are also shown.
    \textbf{c} All four branches shown with the band gap (BG1) as blue shaded area.
    \textbf{d} Enlarged view of the lower two branches.
    The color map of the solid lines represents the intensity of the axial component described by the polarization factor $P_u=\dfrac{m|u_k|^2+m|v_k|^2}{m|u_k|^2+m|v_k|^2 + j|\varphi_k|^2+j|\vartheta_k|^2}$.
    \textbf{e} The linear wave dispersion relationships for negatively twisted lattice (left), natural condition (center, reprinted from panel (\textbf{c})), and positively twisted lattice (right).
    The twist angles for negatively and positively twisted lattices are $-25^\circ$ and $25^\circ$ respectively.
    The Zak phase for each band is labeled in green.
    Trivial and nontrivial band gaps are shaded with pale red and pale blue, respectively.
    The geometrical parameters are: $h_0=30$ mm, $R=36$ mm, $\theta_0^{(1)}=70^\circ$ and  $\theta_0^{(2)}=-70^\circ$.
    }
    \label{fig:dispersion_twist}
\end{figure}

First, we analyze the dispersion relationship of this lattice under natural condition, namely without any applied twist~(i.e., zero twist angle; see Fig.~\ref{fig:dispersion_twist}a).
Figure~\ref{fig:dispersion_twist}c shows the wave dispersion relationship with the unit cell geometrical parameters $h_0=30$ mm, $R=36$ mm, $\theta_0^{(1)}=70^\circ$, and $\theta_0^{(2)}=-70^\circ$~(see {Supplementary Note 2} for the value of $\alpha$ and $\beta$).
% Note that with the given geometrical design where only the initial chirality is set different, we have $\alpha_{11}=\beta_{11}=2.685\times10^{4}$ $\rm N\,m^{-1}$, $\alpha_{22}=\beta_{22}=26.07$ $\rm N\,m\,rad^{-1}$, and $\alpha_{12}=-\beta_{12}=-819.8$ $\rm N\,rad^{-1}$ (see {Supplementary Note 2} for more details).
The color intensity of each branch represents the axial polarization factor $P_u$ of the corresponding eigenvector $\bm{u}_k$~(see Supplementary Note 4 for further discussion).

In this coupled 1D system, we can observe four branches: two lower and two upper branches~(see the enlarged view in Fig.~\ref{fig:dispersion_twist}d; two lower branches are almost collapsing onto each other)~\cite{Yasuda2017a}.
Note that at $k=\pi/h_0$ there is degeneracy between the two lower and the two upper branches.
Between the upper and the lower branches~(i.e., second and third branches), we see a wide band gap denoted as BG1 in Fig.~\ref{fig:dispersion_twist}c.
Notably, this band gap emerges due to only the opposite chirality in our system (i.e., $\alpha_{12}=-\beta_{12}$ in Eq.~\eqref{eq:linear_ansatz}), \textit{without} the necessity of dimerizing axial or rotational stiffness (i.e., without changing any geometric parameters along the lattice to alter $\alpha_{11}$ and $\beta_{11}$). Alternating chirality itself introduces a band gap in the dispersion relation, and this is a key feature of our system distinctive from previous studies~\cite{Kopfler2019,Huda2020}.

%This gap opens because of the difference between two coupling terms, $\alpha_{12}$ and $\beta_{12}$. Opposite chirality of unit cells in our system means $\alpha_{12}=-\beta_{12}$, and therefore, without changing any other geometric parameter along the lattice, alternating chirality itself introduces a band gap in the dispersion relation.

Now, we explore the in-situ tunability of the wave dispersion relationship, without replacing or changing the unit cell design, but by just twisting the lattice.
Here, we assume that the lattice has an even number of unit cells in total~(see Supplementary Note 3 and 5 for more information).
Interestingly, if we quasi-statically twist the lattice along the $z$-axis without changing the total length of the lattice~(i.e., the fixed boundary at both ends), we can change the strain landscape of the lattice.
For instance, starting from the natural configuration~(Fig.~\ref{fig:dispersion_twist}a), as we twist the lattice in the positive $\varphi$ direction, we can see that the negative chirality unit cells~(blue-colored unit) are elongated and the positive chirality unit cells~(red-colored unit) are compressed, as shown in Fig.~\ref{fig:dispersion_twist}b~(see also Supplementary Note 3, Supplementary Figure 1-2 and Supplementary Video 1 for further insights).
Notice that if the total length of the lattice is kept constant, the lattice constant of the supercell (i.e., a pair of unit cells) is also kept constant throughout the twisting process. %regardless of the actual total number of unit cells.

Recall that the unit cell exhibits a nonlinear static response. % regardless of the initial chirality.
Thus the elongated unit cells and the compressed unit cells show different instantaneous stiffnesses, which alters the effective linear stiffnesses $\alpha$ and $\beta$. This, in turn, results in the different linear dynamics of the lattice without replacing the unit cells.
Figure~\ref{fig:dispersion_twist}e shows the linear wave dispersion relationships for three different twist angles per supercell: $\overline{\varphi}_{b}=(-25^\circ,~0^\circ,~25^\circ)$.
In the lower frequency regime, either positively or negatively twisting the lattice induces the first branch to shift slightly downward and the second branch to shift significantly upward.
As a result, the degenerate point at $k=\pi/h_0$ disappears, and the BG1 becomes narrower.
In the higher frequency regime, we see that an additional band gap labeled as band gap 2~(BG2) opens between the third and fourth branches~(see also Supplementary Video 2). Strikingly, the topological nature of these band gaps are highly distinctive, as will be discussed next.
% Note that unlike zero twist angle~($\overline{\varphi}_b=0$), the polarization factor of the first and second branches indicate the coexistence of the axial and rotational component with an approximately equal amount of contribution~(i.e., $P_u\approx0.5$).
We note in passing that the analysis above is specific to the lattice with an even number of unit cells, which shows symmetric behavior when twisted positively and negatively.
See {Supplementary Note 3 and 5}, {Supplementary Figure 1-2} for the comparison between the lattices with even and odd numbers of unit cells.

\subsection{Topological characterization}\label{topology}

For the topological characterization of the system, we first show that the dynamical matrix $\hat{D}_k$ 
can be written as
\begin{equation}
    \hat{P}\hat{D}_{k}\hat{P}^{-1} = \hat{D}_{-k},
\end{equation}
where $\hat{P}$ is the symmetry operator~(please see Supplementary Note 5 for the detail).
% where the symmetry operator $\hat{P}$ reads as
% \begin{equation}
%      \hat{P} = \sigma_0 \otimes \sigma_x = \bcrm 
%      0 & 1 & 0 & 0 \\
%      1 & 0 & 0 & 0 \\
%      0 & 0 & 0 & 1 \\
%      0 & 0 & 1 & 0 \\ \ecrm,
% \end{equation}
% %
% and $\sigma_0$, $\sigma_x$ are the identity and Pauli matrix respectively.
% This symmetry operator, as it was noted in~\cite{Kopfler2019}, is connected with the mirror symmetric contribution of mass and moment of inertia in the unit cell, and results in quantized values ($0$ or $\pi$) of the Zak phase (see also {Supplementary Note 5}).
In the absence of degeneracies, the Zak phase of a band can be obtained by accumulating the phase resulting from the corresponding eigenvectors all over the first Brillouin zone (BZ; Fig.~\ref{fig:dispersion_twist}e).
If we discretize the BZ with $K$ points, the Zak phase for \textit{m}th isolated band is defined as~\cite{Resta2000, Wang2019}
\begin{equation}\label{eq:Zak}
    \phi_{m} = - \sum_{k=0}^{K-1}  \operatorname{Im} \ln \langle \bm{u}_{k}^{m} | \bm{u}_{k+1}^{m} \rangle.
\end{equation}
Here, the band index runs from $m=1$ to $4$.
Except for the case of natural condition (i.e., zero twist angle; see below)
the four bands in the dispersion relation do not have a degenerate point aside from the origin ($\bm{k}=0$).
Therefore, we directly apply Eq.~\eqref{eq:Zak} to obtain the Zak phases of the bands, which are
labeled in the left and right panel of Fig.~\ref{fig:dispersion_twist}e, respectively. 
Indeed the Zak phases of the four bands take the values $0$ or $\pi$, while interestingly enough, these values are switched for the negatively and positively twisted lattices. 
To topologically characterize the band gaps, we sum the Zak phases below the corresponding band gaps~\cite{Xiao2014}. 

We find in Fig.~\ref{fig:dispersion_twist}e that for both negatively and positively twisted lattices, the BG1 holds a sum of Zak phases $\pi$, which indicates that it is a topological nontrivial band gap, a property that does \textit{not} change with the value and sign of the twisting angle.
A special treatment is needed for the case of zero twist angle, because the bands are degenerated at the end of BZ ($\bm{k}=\pi/h_0$)
and therefore Eq.~\eqref{eq:Zak} can not be used.
For such a case, we obtain the topological index for the bands $m$th and ($m+1$)th (that are degenerated at some point)
together via \textit{many band Berry phases} (Wilson-loop eigenvalues)~\cite{Vanderbilt2018, Wang2019}, such that 
\begin{equation}\label{eq:Zak_wilson}
    \phi_{m, m+1} = - \sum_{k=0}^{K-1}  \operatorname{Im} \ln 
    \operatorname{det} \begin{pmatrix}
    \langle \bm{u}_{k}^{m} | \bm{u}_{k+1}^{m} \rangle & \langle \bm{u}_{k}^{m} | \bm{u}_{k+1}^{m+1} \rangle \\
    \langle \bm{u}_{k}^{m+1} | \bm{u}_{k+1}^{m} \rangle & \langle \bm{u}_{k}^{m+1} | \bm{u}_{k+1}^{m+1} \rangle 
\end{pmatrix}.
\end{equation}
In our system, we calculate that $\phi_{1, 2}=\phi_{3, 4}=\pi$ under the natural condition.
We may thus conclude that BG1 is nontrivial even for the case of zero applied twisting. 
%Notably, we mention that this topologically nontrivial band gap emerges due to only the alternating chirality (i.e., $\alpha_{12}=-\beta_{12}$ in Eq.~\eqref{eq:linear_ansatz}), \textit{without} the necessity of dimerizing axial or rotational stiffness (e.g., differing $\alpha_{11}$ and $\beta_{11}$). 
%This is a key feature of our system distinctive from previous studies~\cite{Kopfler2019,Huda2020}.

Now, for BG2, the sum of the Zak phase below BG2 is $\pi$ for the negatively twisted lattices and $0$ for the positively twisted lattices.
This implies that the BG2 exhibit a topological phase transition (from nontrivial to trivial) as the twist angle varies from negative to positive. 
At the special case of zero twist, the gap closes. 
%Such transition can also be understood by the band inversion seen in previous studies~\cite{Xiao2014}.
This distinctive topological nature of the BG2 as a function of the applied twisting angle enables the topological state transfer across the lattice, as will be discussed and verified in the following sections.

\subsection{Emergence of boundary states in finite lattice under fixed boundary conditions}\label{eigenanalysis}

Due to the principle of bulk-boundary correspondence~\cite{Hasan2010}, topologically nontrivial band gaps, identified in the infinite lattice, host topologically-protected boundary states when the lattice is finite.
To study the emergence of boundary states at both BG1 and BG2, we calculate the eigenfrequecies and eigenmodes of our dimer Kresling lattice, as a function of the twisted angle $\overline{\varphi}_b$, for an even number of unit cells~(please see {Supplementary Note 6.2}, {Supplementary Figure 3} for the odd number of unit cell case.).
% To study the emergence of boundary states at both BG1 and BG2, we calculate the eigenfrequecies and eigenmodes of our dimer Kresling lattice, as a function of the twisted angle $\overline{\varphi}_b$, for an even and odd number of unit cells.
Throughout the paper, we employ fixed boundary conditions at both ends
since only this choice can guarantee that the total length of the lattice remains unchanged under twisting and can also keep the lattice twisted at a certain angle.
For instance, if we twist the lattice and release one end such that the boundary condition is fixed-free, then the lattice recoils back to the natural condition where the lattice is not twisted.

\begin{figure*}[htbp]
    \centering
    \includegraphics[width=\linewidth]{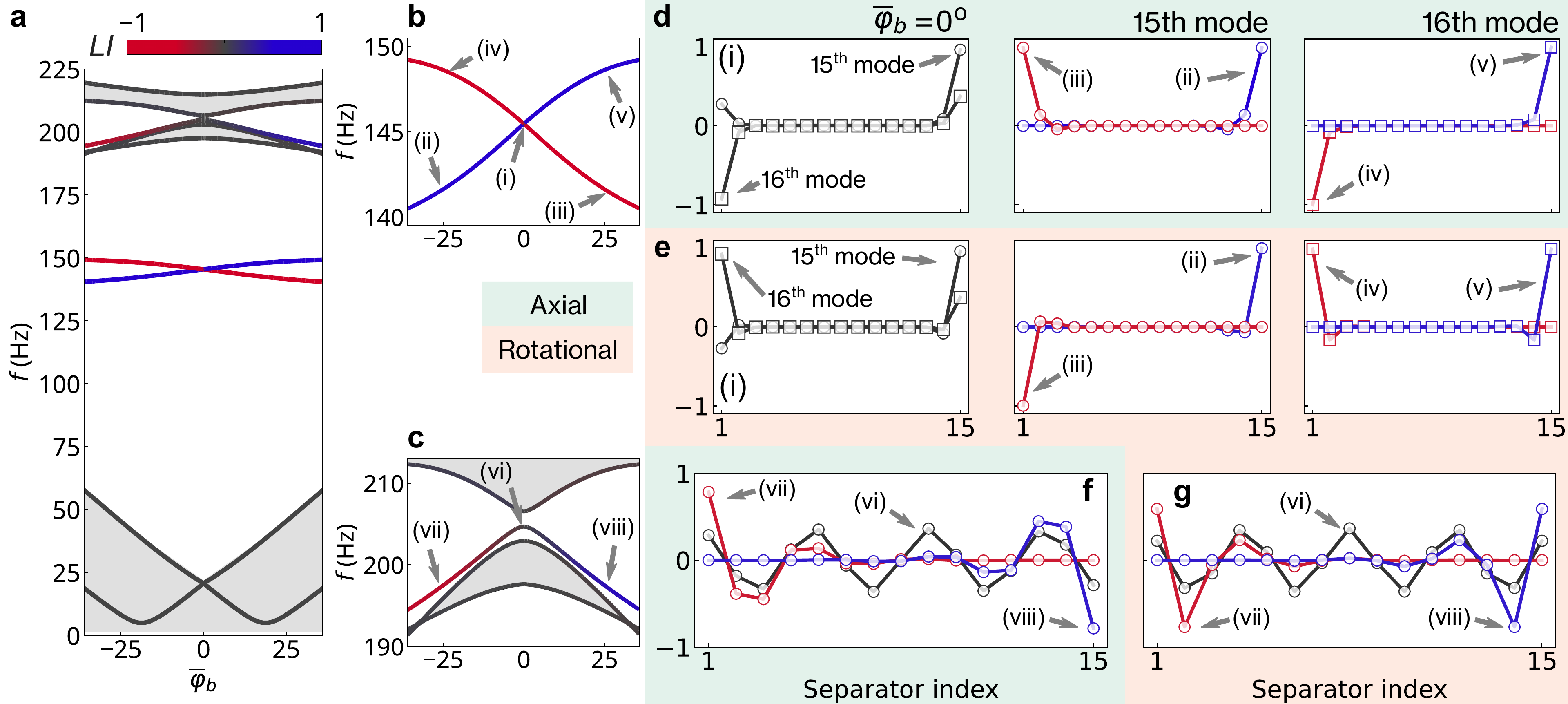}
    \caption{\textbf{Emergence of boundary states in the finite lattice with 16 unit cells under fixed-fixed boundary conditions.}
    \textbf{a} The cut-off frequencies of the four branches in the coupled system, and the boundary state frequencies as a function of twist angle.
    From lowest to highest: upper cut-off frequencies of the first and second branch;
    boundary state mode in BG1 (15th and 16th mode);
    lower and upper cut-off frequencies of the third branch,
    boundary state mode in BG2 (23rd mode);
    lower and upper cut-off frequencies of fourth branch.
    The color intensity of the solid lines represents the localization index $LI$ defined in Eq.~{\eqref{eq:locidx}}.
    \textbf{b}, \textbf{c} show the enlarged view of the boundary state frequencies in BG1 and BG2:
    \textbf{b} 15th and 16th modes; \textbf{c} 23rd mode.
    The areas shaded with the light-gray correspond to the bulk mode frequency regimes.
    \textbf{d}, \textbf{e} The 15th and 16th boundary modes in BG1 in axial (green background panel) and rotational (red) components.
    For the 15th mode, (i) $\overline{\varphi}_{b}=0$, (ii) $-25^\circ$, (iii) $25^\circ$.
    For the 16th mode, (i) $\overline{\varphi}_{b}=0$, (iv) $-25^\circ$, (v) $25^\circ$.
    \textbf{f}, \textbf{g} The 23rd boundary mode in BG2 in axial and rotational components.
    (vi) $\overline{\varphi}_{b}=0$, (vii) $-25^\circ$, (viii) $25^\circ$. 
    }
    \label{fig:eigenmode}
\end{figure*}

%We start with the case of finite lattice with even number of unit cells.
%Recall that this corresponds to odd number of polygonal separators.
Figure~\ref{fig:eigenmode}a summarizes the transition of the cutoff frequencies and boundary state frequencies of the finite lattice with 16 unit cells, as a function of twist angle $\overline{\varphi}_b \in [-30^{\circ}, 30^{\circ}]$.
The color intensity of the lines corresponds to the localization index~(LI) inspired by the inverse participation ratio~(IPR) and the center of mode~(CoM):
\begin{align}\label{eq:locidx}
    LI=IPR\times CoM=\left|\left(K\bm{u}^{\circ2}\right)^{\circ2}\right|\left(\frac{2}{N}\bm{w}^TK\bm{u}^{\circ2}\right),
\end{align}
where $\bm{u}$ is the eigenvector, $K$ is the commutation matrix, and $\bm{w}$ is the weighting vector~(see {Supplementary Note 6.1}).
% where $K$ is the commutation matrix, $\bm{u}$ is the eigenvector, and $\bm{w}=\left(-N/2,\cdots,-1,1,\cdots,N/2\right)^T$ is the weighting vector~(see {Supplementary Note 6.1}).
The power $\circ2$ denotes Hadamard power.
If the eigenmode is skewed toward left~(right), then $LI\to-1$~($LI\to1$).
Note that each entry of $\bm{u}^{\circ2}$ is proportional to the axial and rotational component of the kinetic energy of each polygonal separator.
Therefore, the localization index corresponds to the location of the energy intensity.
To closely examine the boundary states and their eigenmodes, the enlarged views of the boundary states in BG1 and BG2 are shown in Fig.~\ref{fig:eigenmode}b and c, respectively.
Also, their eigenmodes are plotted in both axial and rotational components in Fig.~\ref{fig:eigenmode}d-g.

In BG1, two boundary modes can be observed for all the values of the twisting angle.
This complies well with the fact that BG1 is topologically nontrivial for all the values of twisting (i.e., does not undergo a topological transition as a result of twisting). %, and hence, the system support localized states for all values of twist angles.
At the special case of no twisting, $\overline{\varphi}_b=0$, their frequencies are almost identical and closely degenerate.
Their eigenmodes for this case are shown by label (i) in Fig.~\ref{fig:eigenmode}d and e, where we see non-negligible amplitude at both ends of the lattice. 
When the lattice is twisted positively or negatively, one of the modes shows a decrease in the frequency while the other increases.
If we extract these 15th and 16th modes along these transition curves as denoted in Fig.~{\ref{fig:eigenmode}}b, we can see the highly localized boundary states at either end of the lattice for any $\overline{\varphi}_b$.
Notice that for the 15th mode, the edge state is at the right boundary for the negative twist angle~(label (ii) in Fig.~{\ref{fig:eigenmode}}d), and changes to the left boundary when twisted positively~(label (iii) in Fig.~{\ref{fig:eigenmode}}d).
Similarly, for the 16th mode shown in the right panel of Fig.~{\ref{fig:eigenmode}}d, the edge state changes its localization site from left to right as the twist angle transitions from negative to positive.

In BG2, see Fig.~\ref{fig:eigenmode}c, we observe one boundary mode localized at the left boundary for negative twist angles, as shown in the profiles labeled as (vi) in Fig.~\ref{fig:eigenmode}f and g.
In contrast to the boundary modes of the BG1, this mode is less localized, i.e., it has a smaller localization length, due to the smaller width of the BG2.
Increasing the twisting angle $\overline{\varphi}_b$, the mode becomes more and more extended till we pass $\overline{\varphi}_b=0$, where the band gap becomes extremely narrow for the finite lattice (the band gap closes for the infinite lattice).
For the finite case, at this point the mode transforms into the bulk mode shown in label (vi) in Fig.~\ref{fig:eigenmode}f and g.
Passing to positive twist angles, the mode starts again to be localized, but this time on the right boundary of the lattice, see (viii) in Fig.~\ref{fig:eigenmode}f and g.
Since BG2 undergoes a topological transition with the twist angle, we witness the appearance and disappearance of the edge state for a particular boundary.
The edge state disappears from the left boundary and appears on the right as the twist angle goes from negative to positive.

% When we consider an odd number of unit cells in the origami lattice, namely an even number of polygonal separators, we notice a qualitatively similar behavior (see {Supplementary Note 6.2}, {Supplementary Figure 3}) except that BG2 now hosts two edge states for the negatively twisted configuration and none for the positively twisted configuration.
% This case complies well with the aforementioned finding that BG2 is nontrivial and trivial for negatively and positively twisted configurations, respectively.

% We also note in passing that the emergence of boundary modes in BG2 follows a similar behavior to the ones of the SSH lattice where tuning parameter (in our case the twisting angle) is the ratio of the coupling of the two sublattice particles.
% Indeed, for the case of SSH with \textit{even} number of particles, which corresponds to the odd number of unit cells (i.e., even number of separators) in our system, two boundary modes appears when the band gap is topologically nontrivial and zero when it is trivial.
% In contrast, for the case of \textit{odd} number of particles (i.e., even number of unit cells in our system) always one boundary mode appears which as we tune the coupling of the two sublattice particles, it goes from one end to the other by passing the point where it transforms into a bulk mode (when the ratio of the coupling becomes one, corresponding to the band ``closing'').

\subsection{Experimental demonstration of lower frequency regime dynamic response}\label{lower}

To examine the practicality of our analysis, we now experimentally observe the eigenmodes of the system, especially in the lower frequency regime $f\in(0,180]$ Hz, in which the topological boundary mode frequencies lie within.
The upper frequency regime $f\in(180,220]$ Hz, which hosts the upper brunches and the boundary modes of the BG2, is hindered by excessive amount of dissipation and their experimental observation was not possible.
We construct the lattice consisting of 16 unit cells~(8 positive and 8 negative chirality cells) made of polyethylene terephthalate~(PET) sheet~(see Methods section, {Supplementary Note 7}, and {Supplementary Figure 4-5} for the details).
The initial conditions under consideration are: (i) natural state where $\overline{\varphi}_b=0^\circ$, and (ii) twisted state where $\overline{\varphi}_b=15^\circ$.
The chirp signal with different frequency ranges is applied via a vibration shaker, and the polygonal separator motion is measured and extracted through the digital image correlation technique.

\begin{figure}[htbp]
    \centering
    \includegraphics[width=0.5\linewidth]{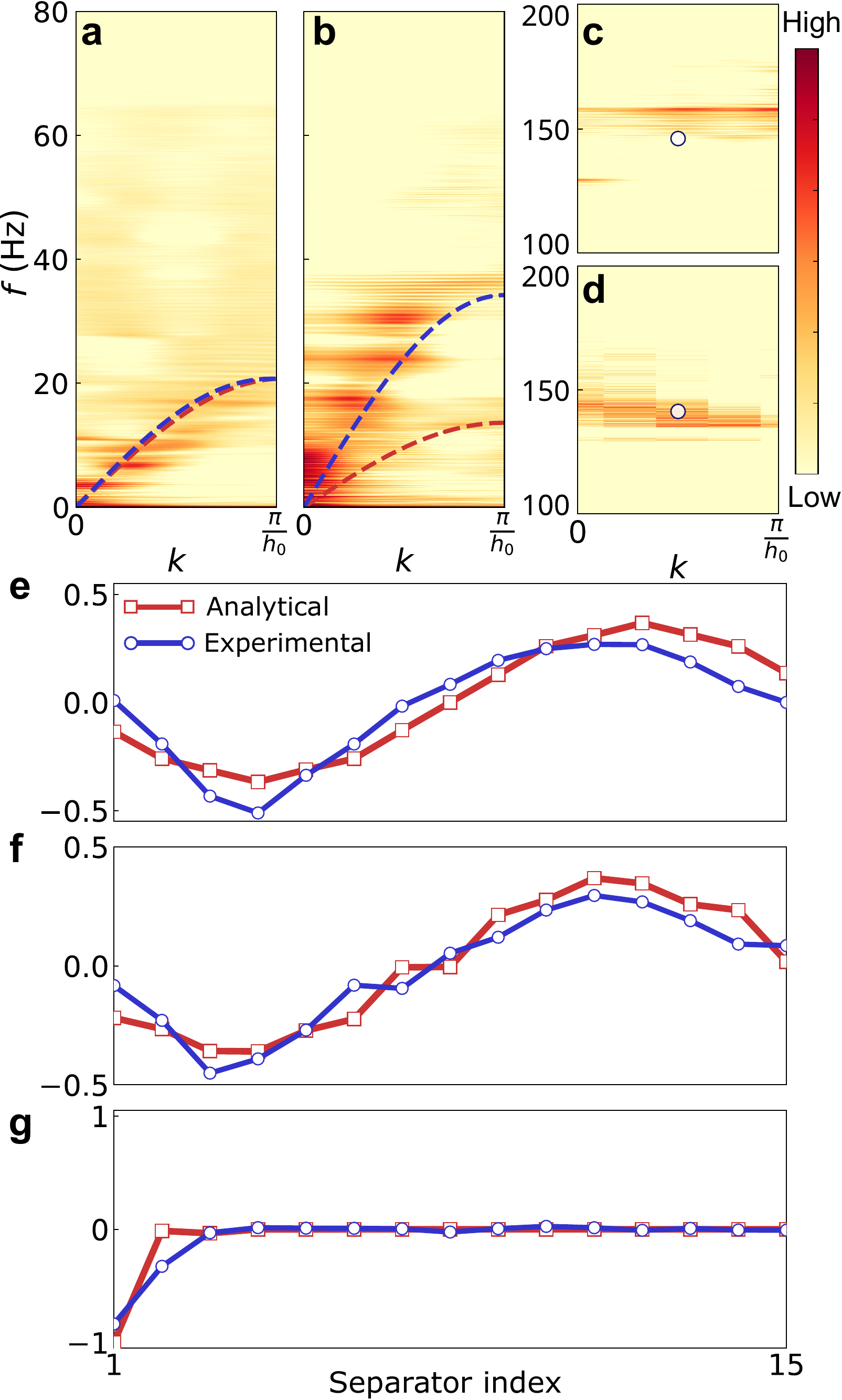}
    \caption{\textbf{Experimentally acquired dispersion relationships from chirp signal response of the 16-unit-cell dimer Kresling origami lattice.}
    The input chirp signal is given for 10 seconds, sweeping the frequency from 0 to 180 Hz.
    For each case, the lattice is quasi-statically twisted for (\textbf{a}, \textbf{c}) $0^\circ$ (natural condition), and (\textbf{b}, \textbf{d}) $15^\circ$ twist.
    Red and blue dashed lines represent the analytically predicted first and second branches, and circle symbol represents the 15th eigenmode (topological mode) within the band-gap.
    The axial displacement component of the normal modes are extracted for both linear eigenanalysis prediction and experiment.
    \textbf{e} Natural condition, the fourth normal mode.
    \textbf{f} Twisted condition, $\overline{\varphi}_{b}=15^\circ$, the eighth normal mode.
    \textbf{g} Natural condition, the 15th mode (topological mode).
%    Red solid lines with square symbol, analytical profile obtained by linear eigenanalysis; blue solid lines with open circle, experiment.
    The experimental normal modes are extracted via SVD-based DMD.
    }
    \label{fig:dispersion_low_exp}
\end{figure}

In Fig.~\ref{fig:dispersion_low_exp}a-d, we show the dispersion relationships extracted by performing the fast Fourier transform (FFT) to the experimental results of the aforementioned two conditions, along with their corresponding analytical predictions. The experimentally obtained dispersion relationships coincide with the analytical predictions denoted as red and blue dashed lines, regardless of the twist angle.
We can clearly see the expansion of the frequency regime covered by the lower two dispersion curves, by comparing Fig.~\ref{fig:dispersion_low_exp}a and b.
Figure~\ref{fig:dispersion_low_exp}c-d show the FFT result of the excitation within BG1, specifically at $f\in[100,\,200]$ Hz.
First, at the zero twist angle, we can clearly see high intensity region near $f=158$ Hz~(Fig.~\ref{fig:dispersion_low_exp}c), which is slightly higher than the eigenanalysis prediction at $f\approx147$ Hz, denoted as open circle.
Similarly, at $\overline{\varphi}_b=15^\circ$ shown in Fig.~\ref{fig:dispersion_low_exp}d, the high intensity region of FFT lies near $f=142$ Hz.
The eigenanalysis in this case predicts the mode frequency to be $f\approx144$ Hz, which slightly overestimates.
Although there are minute disparities in boundary mode frequencies between analytical prediction and experimental results, these FFT results still suggests the clear mode excitation within the band gap.
Moreover, the decreasing behavior of the boundary mode frequency as a function of twist angle is present in both eigenanalysis and the experiment.

For closer examination, we extract the axial displacement component of the normal mode with an aid of dynamical mode decomposition~(DMD) based on the singular value decomposition~(SVD)~\cite{Kutz2016}.
First, we extract the fourth mode of the natural condition in Fig.~\ref{fig:dispersion_low_exp}e.
When the lattice is under natural condition, the experimental results (circle symbol with blue solid lines) qualitatively follows the profile of analytical prediction~(square symbol with red solid lines), except for the spatial decay of the amplitude toward the right end of the lattice.
We believe that this is due to the energy dissipation in the experiment, where the lattice is excited at the left end~(see {Supplementary Note 8}, {Supplementary Figure 6} for a numerical investigation with a dash-pot model).
Similarly, in Fig.~\ref{fig:dispersion_low_exp}f, the experimental (analytical) profile of the eighth mode under the twist of $\overline{\varphi}_b=15^\circ$ is shown in blue (red) line, showing the agreement between experimental and analytical results.

Within the band gap, linear analysis predicts the localization at the left end for the 15th mode~(red line in Fig.~\ref{fig:dispersion_low_exp}g).
If we extract the normal mode near $f=158$ Hz from the experimental result~(where FFT shows high intensity in Fig.~\ref{fig:dispersion_low_exp}c), we see the equivalent localization phenomenon as the blue solid line in Fig.~\ref{fig:dispersion_low_exp}g shows.
Recall that our eigenanalysis in the previous sections identified the mode inside the BG1 as topological mode regardless of the twist angle.
This implies that the experimentally observed normal mode in Fig.~{\ref{fig:dispersion_low_exp}}g is a topological mode, which is induced solely by dimerizing the coupled nature.
For other normal modes extracted from experimental results, see {Supplementary Note 8} and {Supplementary Figure 6}.

\subsection{Numerical investigation of topological state transfer}\label{upper}

Now, we explore the evolution of the topologically localized modes in BG1 and BG2 as the lattice is twisted over time.
Inspired by the works of state transfer in odd-sized quantum~\cite{Mei2018} and classical~\cite{Brouzos2020} SSH lattices, and the similar behavior of the boundary modes in BG2, we consider the case of odd number of separators (even number of unit cells).
%We also note in passing that this investigation relies on numerical simulations, since the experimental observation of high frequency modes (BG2) was hindered by excessive amount of dissipation.
The overall objective here is to explore the possibility of the in-situ boundary state transfer from one end of the lattice to the other as we apply a twisting on the lattice. 
For that, we numerically solve the fully nonlinear equations of motion of our dimer origami system~(Eq.~{\eqref{eq:eom}}) consisting of 32 unit cells, pre-rotated for the total twist angle of $\varphi_b=-240^\circ$~(equivalent to the supercell twist angle $\overline{\varphi}_b=-15^\circ$).
The protocol we follow consists of two phases: (1) the excitation and (2) loading phase.
First, we apply at the left boundary a sinusoidal signal to excite the boundary mode in BG1 (BG2), which is located at $f=147.8$ Hz ($200.4$ Hz).
This excitation phase lasts 10 seconds~(i.e., $t\in[-10,0]$).
In the loading phase, we halt the boundary driving signal and gradually twist the lattice at the assigned twist rate until the twist angle reaches $\varphi_b=240^\circ$~($\overline{\varphi}_b=15^\circ$).
(See {Supplementary Table 1} for the detail of the numerical values.)
We employ loading protocol that initiates and terminates smoothly, to avoid the excitation of other modes.
The boundary angle profile $\varphi_b$ is expressed as,
\begin{align}\label{eq:protocol_sin}
    \varphi_{b,sin}(t)&=\varphi_b^{(0)}+\left(\varphi_b^{(1)}-\varphi_b^{(0)}\right)\sin^2\left(\frac{\pi t}{2T}\right),
\end{align}
where $T$ is the total loading time, and the superscripts $(0)$ and $(1)$ refer to the initial and final state.
Note that $C^1$ continuity at $\varphi_b=\varphi_b^{(0)}$ and $\varphi_b=\varphi_b^{(1)}$ guarantees the smooth initiation and termination of the loading.
The loading phase, and thus the function Eq.~{\eqref{eq:protocol_sin}}, is defined in the time domain $t\in[0,T]$.
Generally, we employ large loading times $T$ ($T=80$ s in this study) to guarantee the quasi-adiabatic evolution which is necessary to avoid the excitation of other modes during the process.
%As a rule of thumb, the value of the $T$ should be in the order of the inverse of the minimum eigenfrequency distance between the ones of the boundary and the bulk modes, all along the process. 

\begin{figure*}[htbp]
    \centering
    \includegraphics[width=\linewidth]{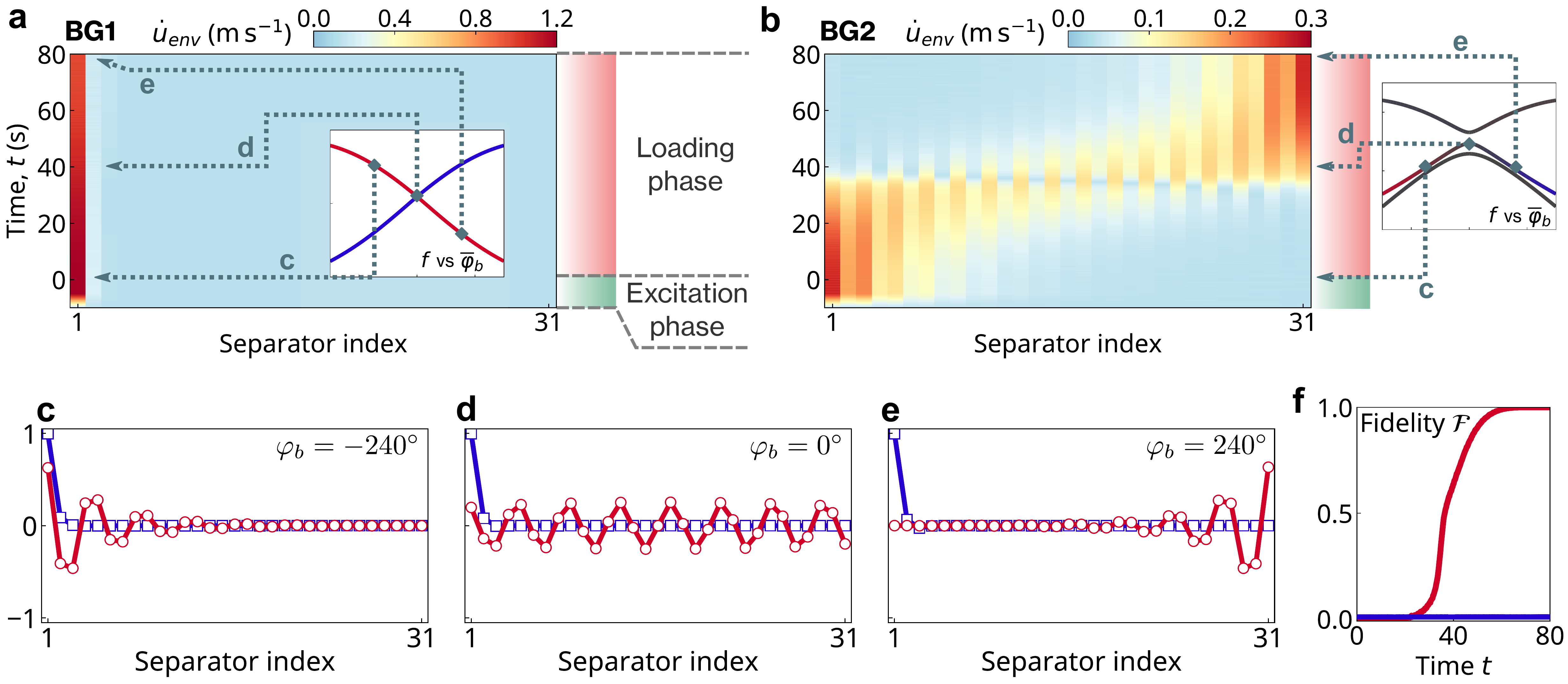}
    \caption{\textbf{Numerical calculation of the topological state transfer through the dimer Kresling lattice.}
    Axial velocity envelope fields $\dot{\bm{u}}_{env}(t)$ for the case of \textbf{a} BG1 and \textbf{b} BG2.
    Unit is $\rm m\,s^{-1}$.
    The first 10 seconds~($t\in[-10,0]$ s) of the numerical solution is the excitation phase at 147.8 and 200.4 Hz respectively.
    Within $t\in[0,\,80]$ s~(therefore $T=80$ s), the lattice is quasi-statically twisted in $\varphi_{b}\in[-240^\circ,\,240^\circ]$.
    The subplots schematically illustrate the transition of excited eigenmodes as a function of twist angle~(similar to Fig.~\ref{fig:eigenmode}b and c, but with 32 unit cells).
    Grey arrows point toward the corresponding instances in the numerical simulations.
    The extracted normal modes of the axial velocity profiles during the quasi-static loading phase are plotted for \textbf{c} $-240^\circ$, \textbf{d} $0^\circ$, and \textbf{e} $240^\circ$.
    Red solid lines with open circle symbol, BG2; blue solid lines with open square symbol, BG1.
    \textbf{f} The time evolution of the fidelity $\mathcal{F}$ defined by Eq.~\eqref{eq:fidelity}.
    }
    \label{fig:energytransfer}
\end{figure*}

Figure~{\ref{fig:energytransfer}}a and b show the numerically solved velocity field in axial component (see {Supplementary Note 9}, {Supplementary Figure 7} for rotational component) for BG1 and BG2 respectively. 
For a better visibility, we numerically estimate the envelope function of the velocity field using the Hilbert transform~(see also {Supplementary Note 9}).
For the case of BG1~(Fig.~{\ref{fig:energytransfer}}a), only the first polygonal separator shows the large amplitude, while the motions of the other separators are negligibly small regardless of the twist angle.
For the case of BG2~(Fig.~{\ref{fig:energytransfer}}b), however, we observe the interesting phenomenon of the energy transfer through the lattice, showing the high intensity region traveling from the left to the right end of the lattice as we gradually apply the rotation.

If we compare the mode shapes extracted at $t=0,\,40$, and $80$ s (corresponding to ${\varphi}_b=-240^\circ,~0^\circ$, and $240^\circ$), we can evidently see the different behavior of the energy management between BG1 and BG2. 
For the extracted mode shapes of BG1 (denoted by blue solid lines with open squares in Fig.~\ref{fig:energytransfer}c-e), we always observe the localization of energy on the left end.
This is in agreement with the previous analysis in Fig.~\ref{fig:eigenmode} (also reprinted in the inset panel of Fig.~\ref{fig:energytransfer}a) that all these modes are sitting on the eigenmode transition curve in the consistent red color, which represents the preservation of the localization index $LI$ (i.e., localization in the left end).

In the case of BG2, however, we observe the transfer of the edge mode from the left to the right boundary, as shown in Fig.~\ref{fig:energytransfer}c-e~(denoted as red solid lines with open circle).
At ${\varphi}_b=-240^\circ$ and $240^\circ$, the boundary states clearly show the localization on the left and right end of the lattice, respectively, which decays exponentially as presented in Fig.~\ref{fig:energytransfer}c and e.
As noted before, during the transition between these two angles, the normal mode can exhibit sinusoidal-like bulk mode profile~(Fig.~\ref{fig:energytransfer}d) when quasi-static torsion is absent (${\varphi}_b=0^\circ$).
This normal mode is reminiscent of those shown through linear eigenanalysis in Fig.~\ref{fig:eigenmode}f, and it is the mode that plays as a vehicle to transfer the energy through the lattice between the two localized modes.
This can be confirmed by the eigenmode transition curve in the sub-panel of Fig.~\ref{fig:energytransfer}b, where we evidently see the change of the color---from red to blue, representing the migration of the localization index $LI$---along the curve.
(See also Supplementary Video 3 for the comparison between BG1 and BG2 cases.)

To quantify and confirm the state transfer from one edge to another, we define the fidelity $\mathcal{F}$ via the dot product~\cite{Caneva2009,Mei2018},
\begin{align}\label{eq:fidelity}
    \mathcal{F}=|\bm{u}_{target}\cdot\bm{u}(t)|,
\end{align}
where $\bm{u}=(u_1,v_1,\varphi_1,\vartheta_1,\cdots,u_N,v_N,\varphi_N,\vartheta_N)^T$ is the normal mode vector based on the envelope function extracted through Hilbert transform~(see {Supplementary Note 9}).
The subscript $target$ refers to the analytically predicted normal mode at $\varphi_b=240^\circ$, which corresponds to the excited eigenfrequency.
Note that when the numerically solved normal mode vector $\bm{u}$ is similar to the analytical prediction $\bm{u}_{target}$, we obtain $\mathcal{F}\to1$.

Figure~\ref{fig:energytransfer}f shows the fidelity as a function of time for the cases shown in Fig.~\ref{fig:energytransfer}a and b in the loading phase, $t\in[0,\,80]$ s.
In terms of transfer fidelity, we can even clearly see that the boundary state in BG2 transfers robustly and efficiently from left to right~($\mathcal{F}=0$ to $\mathcal{F}=1$, denoted as red solid line).
The fidelity smoothly increases and saturates to $\mathcal{F}\approx1$.
% The fidelity gradually starts to rise around $t\approx25$ s, followed by the abrupt increase reaching $\mathcal{F}=0.5$ at $t\approx40$ s.
% In the latter half of the loading phase~($t\in[40,\,80]$ s), the fidelity soon saturates to $\mathcal{F}\approx1$.
In contrast, as suggested by the velocity field shown in Fig.~\ref{fig:energytransfer}a, the fidelity for the case of BG1 is always zero throughout the simulation.
Further investigations on the fidelity in relation with the loading protocol and total loading time can be found in {Supplementary Note 9}, {Supplementary Figure 8}.

This numerical investigation shows highly distinctive characteristics of energy management between BG1 and BG2.
Unlike the boundary states in BG1, the emerging boundary state in BG2 shows the continuous transition from localized and bulk mode profiles, which in turn enables the topological state transfer of the intended boundary mode. This is consistent with the finding in Section D that BG1 is topologically robust without changing its nature, while BG2 shows topological transition from non-trivial to trivial when the lattice is twisted.
We thus verify two distinctive wave dynamics phenomena that can be achieved in the same setting of the versatile Kresling origami platform: robust preservation vs. transfer of the energy localization, simply by changing the boundary condition. 

\section{Discussion}\label{conclusion}
In conclusion, we have analytically, numerically, and experimentally explored the wave dynamics of the dimer Kresling lattice in relevance to the tunable wave dispersion relationships and topological boundary state.
By stacking two origami unit cells of opposite chirality along a lattice, a topological band gap (BG1) emerges in the dispersion diagram.
We experimentally verify the existence of topological edge modes inside the band gap.
Furthermore, under quasi-static torsion given to the lattice, each unit cell exhibits the variable effective linear stiffness, which results in the tunable linear wave dispersion relationships.
Such in-situ tuning opens an additional band gap (BG2) in the dispersion diagram.
While BG1 always stays open and topologically non-trivial, BG2 transitions from a topologically nontrivial to a trivial regime only by changing the twist in the lattice from negative to positive.
As a consequence, a finite length lattice shows the appearance and disappearance of an edge state from the boundary.
We have utilized this fact to facilitate the topological state transfer by varying the twist angle in time.
The transfer fidelity has shown to be extremely high, especially for the loading protocol initiating smoothly and relatively in a gradual manner.
Interestingly, while efficient state transfer has shown in BG2, the boundary state in BG1 stays on one end even under the quasi-static torsion.
Consequently, by the virtue of rich topological characteristics, we have demonstrated two distinct wave manipulation strategies in single design of 1D lattice: highly robust energy confinement capability in BG1, and the highly efficient energy transfer in BG2. This has been achieved simply by changing the boundary condition, without altering pre-configuration of the system or employing complex active mechanisms.
Therefore, our system combines the concept of topology and origami to offer a unique way to harness multi degree-of-freedom of an elastic metamaterial for the purpose of extremely versatile and controllable energy management.

%TC:ignore
\section*{Methods}
\subsection{Fabrication and Experiment}
The Kresling unit cells are fabricated with a polyethylene terephthalate~(PET) sheet of 0.254 mm thickness, cut into the petal-like folding pattern~(see Fig.~S) by the laser cutting machine~(Universal Laser Systems, Inc. VLS4.60).
In this study, we prepare the Kresling unit cells with positive and negative chirality.
Every creaseline is replaced with the compliant mechanism with the discorectanglular shape perforation to accurately fold the creaseline.
We then assemble the dimer lattice of 16 unit cells, with spherical markers attached to the vertices of hexagonal separators.
Note that the markers facing the front side are colored green to facilitate motion tracking.
The assembled lattice is connected to the shaker~(B\"uel \& Kj{\ae}r LDS V406 Shaker) controlled via function generator~(Keysight Technologies, Inc., Agilient 33220A Function / Arbitrary Waveform Generator).
The lattice under vibration is then recorded with two high-speed cameras~(Kron Technologies, Inc., Chronos 1.4).
The recorded videos are post-processed via digital image correlation to extract the time series data of each marker.
For more detail and visual representation, see {Supplementary Note 7, Supplementary Figure 4 and 5}.

\section*{Data Availability}
Data supporting the findings of this study are available from the corresponding author on request.

\section*{Code Availability}
Computer code written and used in the analysis is available from the corresponding author as per requested.

\section*{Author contributions}
Y.M. and J.Y. conceived the project.
Y.M. conducted and carried out the numerical simulations.
Y.M., C.-W.C., and R.C. explored the theoretical framework and prepared the manuscript.
Y.M., C.-W.C., T.S.G., and G.Y. performed the experiment and analyzed the data.
All authors extensively contributed to the work and finalizing the manuscript.
G.T. and J.Y. supervised the project.

\section*{Competing interests}
The authors declare no competing interests.

\begin{acknowledgements}
We acknowledge financial support from the US National Science Foundation under Grant No.~CAREER-1553202 and CMMI-1933729.
Y.M. and J.Y. are grateful for the support of the Washington Research Foundation.
R.C. acknowledges the Startup Grant provided by the Indian Institute of Science.
\end{acknowledgements}

\bibliographystyle{apsrev4-2}
\bibliography{topology}

% ==============================================================
%   Figure 
% ==============================================================

%TC:endignore

\end{document}